\documentclass[aps,prl,
 amsmath,amssymb]{revtex4-1}
\usepackage{graphicx}% Include figure files
\usepackage{bm}% bold math
\usepackage[usenames, dvipsnames]{color}

\usepackage[utf8]{inputenc}
\usepackage[T1]{fontenc}
\usepackage{mathptmx}
\usepackage{hyperref}
\usepackage{caption,setspace}
\usepackage{multirow}

\begin{document}

\title{A general classification scheme of detecting spatial and dynamical heterogeneities in super-cooled liquids}

\author{Viet Nguyen}
\author{Xueyu Song}
\email{xsong@iastate.edu}
\affiliation{Ames Laboratory and Department of Chemistry, Iowa State University, Ames, IA, USA}
\date{\today}

\begin{abstract}
A computational approach via implementation of the Principle Component Analysis (PCA) and Gaussian Mixture (GM) clustering methods from Machine Learning (ML) algorithms to identify domain structures of  supercooled liquids is developed. Raw features data are collected from the coordination numbers of particles smoothed using its radial distribution function and are used as an order-parameter of disordered structures for GM clustering after dimensionality reduction from the PCA. To transfer the knowledge from features(structural) space to configurational space, another GM clustering is performed using the Cartesian coordinates as an order-parameter with the particles' identity from GM in the feature space. Both GM clustering are performed iteratively until convergence. Results show the appearance of aggregated clusters of nano-domains over sufficient long timescale both in structural and configurational spaces with heterogeneous dynamics.  More importantly, consistent nano-domains tilling up the whole space regardless of the system size are observed and our approach can be applied to any disordered systems.
\end{abstract}
\maketitle

Understanding the physics of supercooled liquids near glassy transition remains one of the major challenges in condensed matter science. There has been long recognized that both dynamical and spatial domains in supercooled liquids are heterogeneous~\cite{Cubuk,Walter_Kob,WGotze,STILLINGER,Sastry}. As liquid is cooled far below its melting point fast, dynamics in some regions of the sample can be orders of magnitude faster than the dynamics in other regions only a few nanometers away. However, to identify such domain structures both structurally and configurationally in a consistent fashion and the connection between structures and dynamics remains elusive.

Even though the radial distribution function, {\it g(r)}, has been commonly used to describe the structure of disordered systems such as supercooled liquids for a long time,  but this highly aggregated representation of the whole system could not provide a  realistic description of the spatial heterogeneity. If the coordination number (CN) of a particle in the system  is used to capture its local structure, hence the number of such local structures will be the same as the number of particles in the system.  For a particular configuration of the system, for example, a snapshot from a molecular simulation of the supercooled system or an image of supercooled colloidal system from confocal microscopy,  a useful structural description of the system will be to classify these local structures into a few meso-states, which represent the overall structural heterogeneity of the system. Then the particles in the same meso-state should form domains in the configurational space, and these domains from various meso-states can tile up the whole system, which provide a classification scheme both structurally and configurationally. Furthermore, the meso-states in the structural space and domains in the configurational space should have long lifetimes
to make further analysis possible. For example, dynamical  heterogeneity can manifest itself by monitoring the distribution of diffusion constants of particles in various meso-states. 

Motivated by the necessity of such a classification scheme in supercooled liquids, a computational approach via ML algorithms such as PCA and K-means clustering ~\cite{BishopChristopherM2006Pram,F.Noe-pyemma,G.James} is developed to demonstrate that the structures of supercooled liquids can be classified into a few structurally distinct nano-domains that tile up the whole configurational space with long lifetimes and dynamically differ from each other by calculating diffusion constant distributions from the mean square displacements (MSD). 

\begin{figure}
\parbox{3.3in}
{
		\includegraphics[width=3.0in]{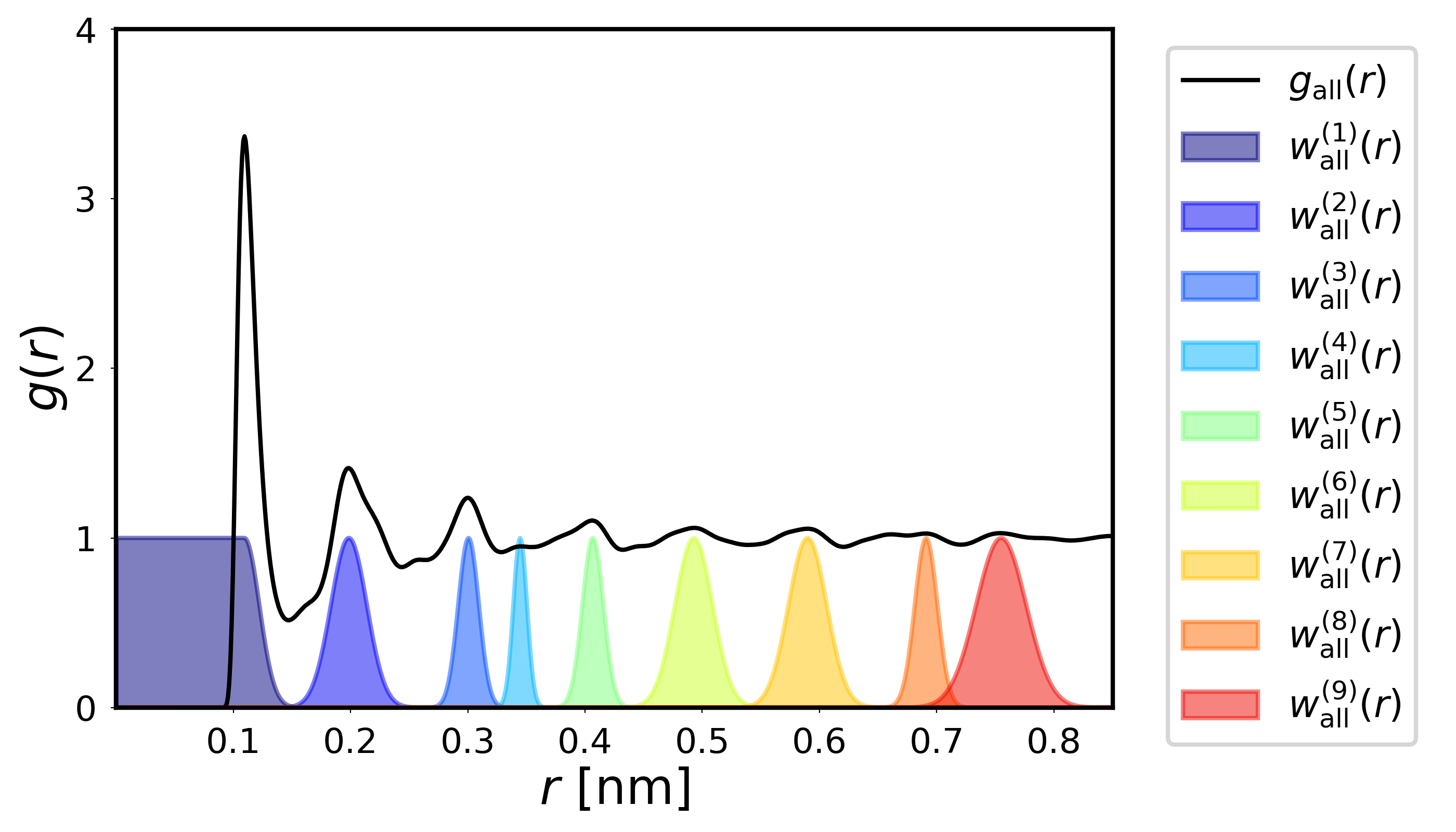}
}
\parbox{3.3in}
{
		\caption{Radial distribution function (rdf) employed by weighted Gaussian functions for Broughton-Gilmer LJ system~\cite{ Gilmer-0001-61608390112-8}at $T^*=0.58$ regardless of particles identities. The color and superscript indicate the WCN's position and number along the rdf.} 
		\label{fig:wcn}
}
\end{figure}

To illustrate such a classification scheme, a well studied system with known classification results is used to demonstrate the idea. Homogeneous liquid-solid nucleation serves as a good example to test our classification scheme by reproducing the known identity of co-existing liquid and solid-like particles. For example, Lechner and Dellago~\cite{Lechner_1.2977970} have shown that, via order parameter $\overline{q_4q_6}$, the identity of liquid and solid particles can be obtained. To make a direct comparison with our classification scheme, an equilibrated Lennard-Jones(LJ) system~\cite{Vega,Gilmer-0001-61608390112-8,Gunawardana-1.5021944,Laird-1.3231693,X.Bai_1.2085147} with a spherical solid (FCC in this case) cluster of a certain size in a supercooled liquid
is generated.  

Using molecular dynamics simulations of this model system, the CN of a particle can be calculated. However, CN-based features suffer a strict cut-off value to determine whether a neighbor particle is counted as in or out of the shell. To avoid this hard assignment, weighted coordination numbers (WCNs)~\cite{rudzinski_1.5064808}, which utilize the normalized Gaussian distribution based on the shell structure of the system {\it g(r)}(Fig.~\ref{fig:wcn}) to weight the contribution of each surrounding particle  based on the particles distance to the central particle. Using the solvation shell features as maxima and minima along the radial direction, the normalized Gaussian distribution functions are placed at the center of these shell features as shown in Fig.~\ref{fig:wcn}. The width of the Gaussian functions depends on the area that the solvation feature covers and neighboring Gaussians such that the value of the intersection is either 0 or roughly 0.25. The WCNs smooth out transitions between solvation shells by counting the particles sitting at the center of the features as one while the one further away from the center feature is counted as a fraction based on the Gaussian distribution function. For each configuration, employing this WCN implementation, each component of a particle's WCNs vector is determined by summing the weight from all surrounding particles within that shell and the dimension of the WCN vector is determined by the number of shells reasonably covers the main features of the {\it g(r)}, $N=9$ in Fig. \ref{fig:wcn}, other numbers of shells tested yield consistent results.

For a single configuration of the simulation, WCNs of all particles are collected from the particles' coordination numbers smoothed using the {\it g(r)}, which form the feature space of our PCA. In general, the features data for the entire system at a particular configuration is represented by a matrix of {\it M{\rm x}N} which is obtained from {\it N} WCNs in a {\it M} particles system in one configuration. The covariance matrix of {\it N{\rm x}N} can be computed  and then diagonalized to form a basis ~\cite{J.Shlens} in the WCN feature space(called PC-space). The inner product of a particle's WCNs with the basis provides a representation of the particle's local structure(called PC representation), K-means clustering shows that in such a representation the particles can be classified into a few clusters, where each such a cluster is a meso-state. Furthermore, as we know the identity of the particles, each meso-state forms a domain in the configurational space. 

Although K-means method is able to cluster particle's PC representation in its PC-space, a major drawback of the K-means algorithm is the lack of an optimally pre-determined number of clusters. Elbow test is used to determine the optimal number of clusters. In this case, the Elbow test correctly predicts K=2,  which is the optimal number of clusters without any knowledge of the system.

\begin{figure}
\resizebox{0.8\columnwidth}{!}
{
\includegraphics[width=1.5in]{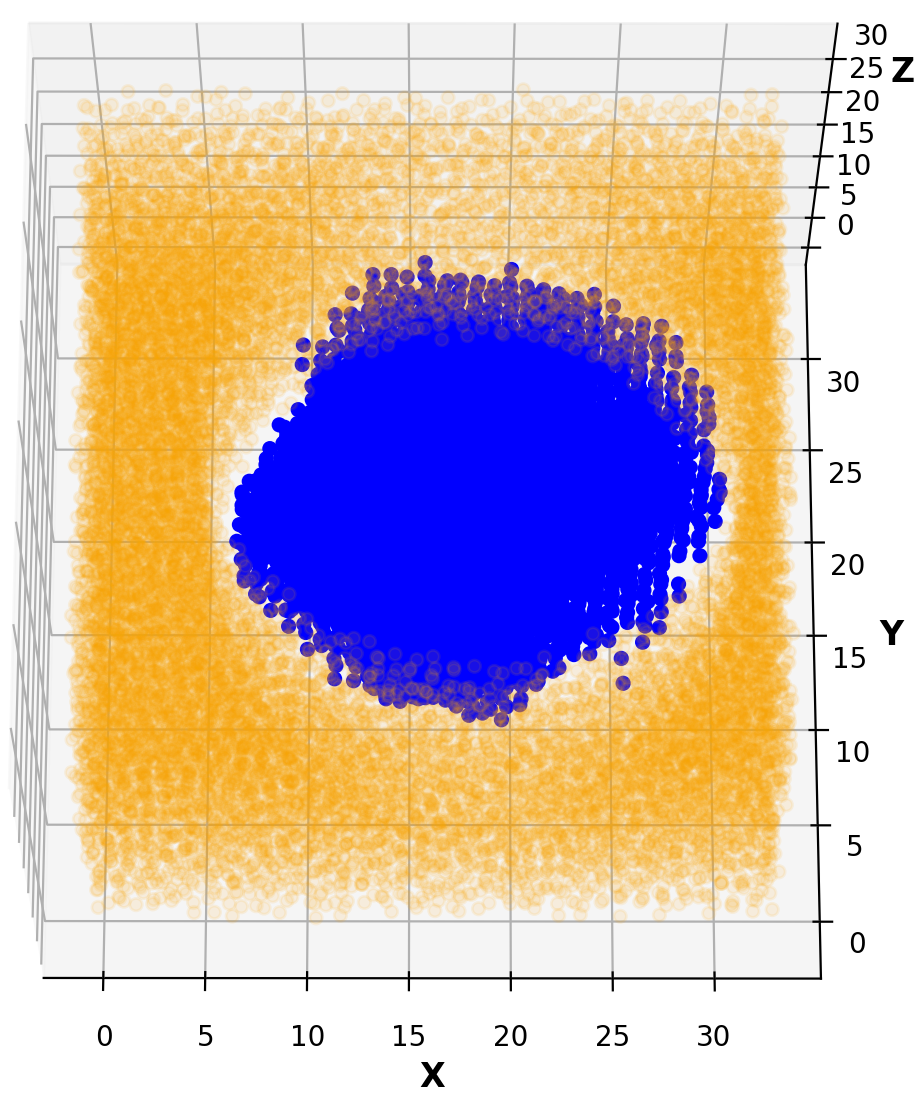}
\includegraphics[width=1.5in]{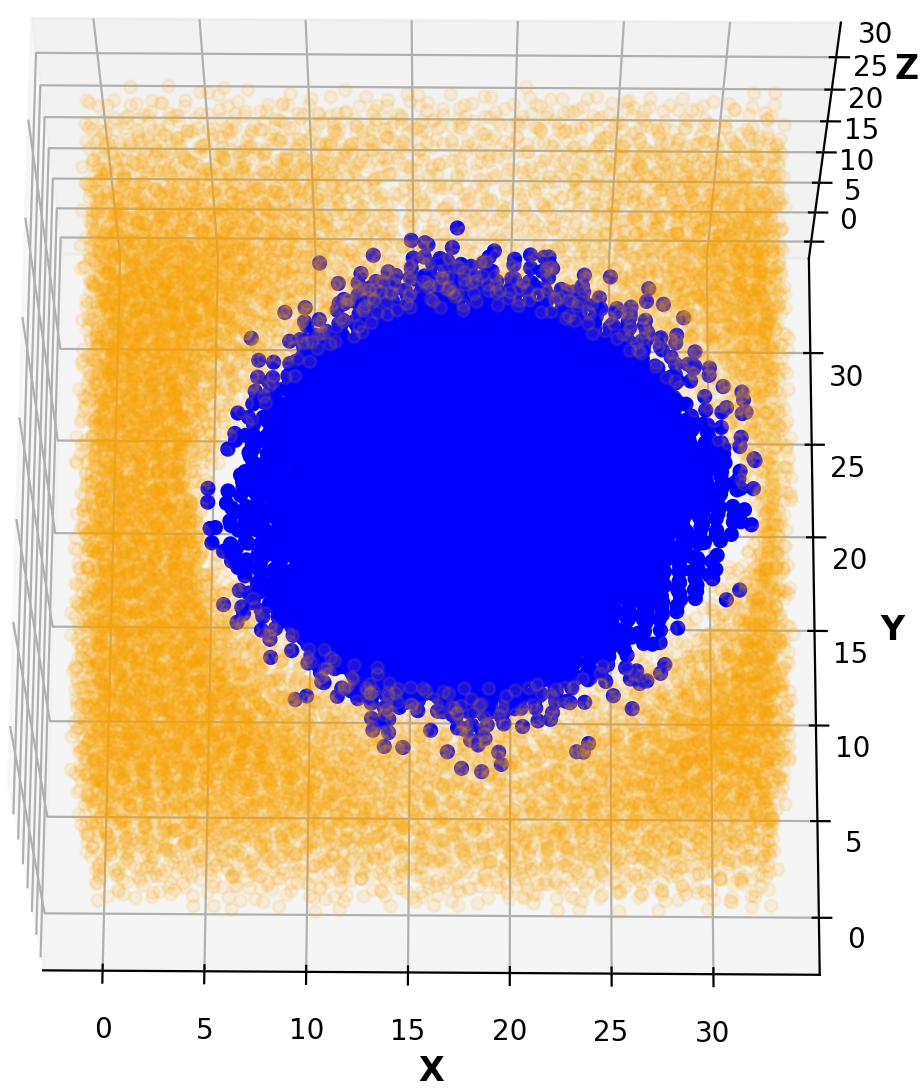}
}
\resizebox{0.8\columnwidth}{!}
{
\includegraphics[width=1.5in]{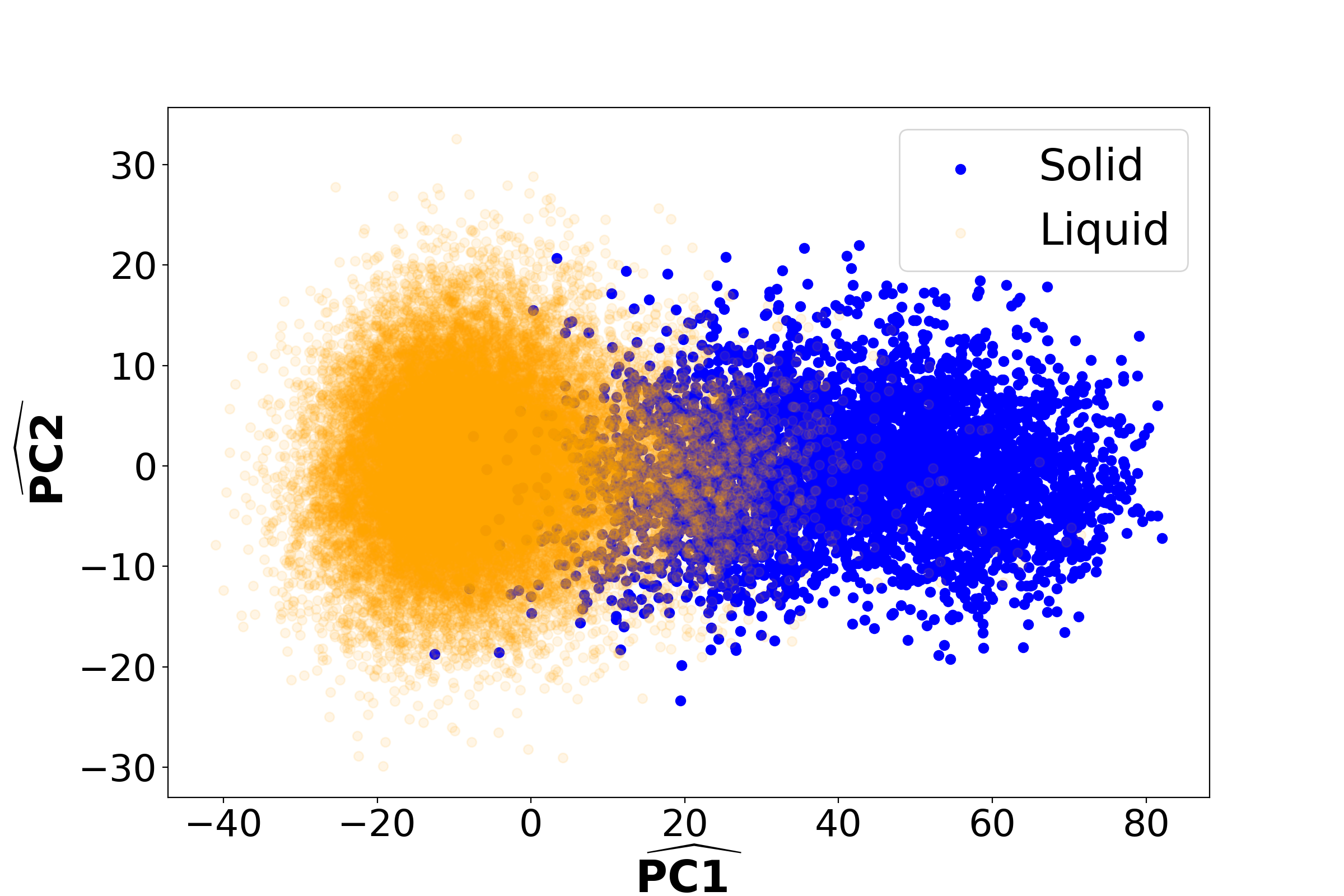}
\includegraphics[width=1.5in]{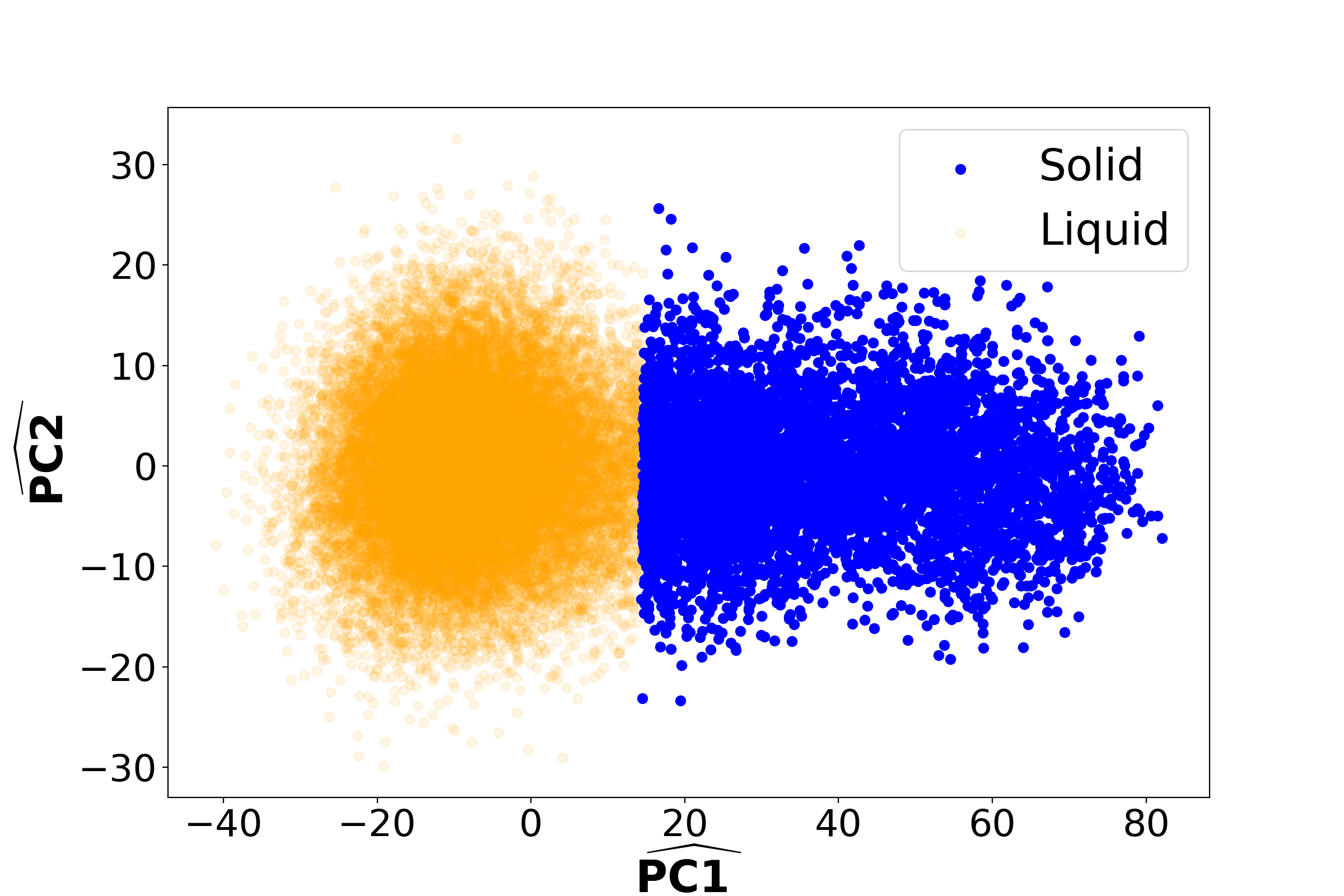}
}
\parbox{6.3in}
{
\caption{ a) Classification of solid-liquid particles using $\overline{q_4q_6}$ classification~\cite{Lechner_1.2977970}.
 b) K-Means clustering using the PC-space information in panel d) without prior knowledge of $\overline{q_4q_6}$.   
 c) PC-space clustering reconstructed from the information of panel a).
 d) K-means clustering using PCA using our scheme without prior knowledge of $\overline{q_4q_6}$ . 
 Solid and liquid-like particles are shown in blue and orange respectively. 
 The hat symbol for labelling axes of PC space represents the inner product of a particle's WCNs with the PCs basis. 
 The clustering is in the full PC-space (9 in this case), but the figure only presents the first two dimensions.
 In the panels a) and b), some of liquid particles covering the solid particles along the Z direction are removed to show the solid cluster clearly.}
\label{fig:nucleation}
}
\end{figure}

Given K=2, Fig. \ref{fig:nucleation} presents a direct comparison between the classification based on the well known $\overline{q_4q_6}$ criterion both in PC-space and configurational space. Panel a) of  Fig. \ref{fig:nucleation} shows the liquid and solid particles classification using $\overline{q_4q_6}$ criterion.
Panel c) of  Fig. \ref{fig:nucleation} shows the identity of particles in the PCs space using the $\overline{q_4q_6}$ criterion as once the classification is done using $\overline{q_4q_6}$, the particles' clustering in the PC-space can be obtained using their associated identities. 

Panel d) of Fig. \ref{fig:nucleation} presents the particles clustering  in PC-space from our PCA and K-means scheme without any knowledge of the $\overline{q_4q_6}$ classification. The corresponding particles clustering in configurational space using the information from the PC-space of panel d) is shown in panel b) of the figure. It is found that the results of distinguishing solid and liquid particles by the means of PCA and K-means scheme are stunningly accurate in comparison with the conventional $\overline{q_4q_6}$ classification. The minor difference between the two methods is the particles in the interfacial region which can be improved using soft clustering algorithms as discussed  below.

To demonstrate the capability of our classification scheme for disordered systems, a well studied binary LJ model system, the Kob-Andersen model~\cite{kob_andersen_PhysRevE.51.4626,Thomas,WalterKob-scaling}, is used since it is known 
that the model system does not crystallize when it is supercooled well below the melting temperature. Both type of the particles (80\% A and 20\%B) interact via LJ potentials with $\epsilon_{AA}=1$, $\sigma_{AA}=1$, $\epsilon_{AB}=1$ , $\sigma_{AB}=0.8$,$\epsilon_{BB}= 0.5$ and  $\sigma_{BB}=0.88$ with the conventional meaning of the LJ potential and  the masses being 1 for both type of particles. The cut-off distance is 2.5$\sigma$ and periodic boundary conditions are used.  Using LAMMPS~\cite{LAMMPS}, Nose-Hoover thermostats are applied for controlling the pressure and temperature of the system. The external pressure is set to zero for all simulations. The time unit for the LJ system is the reduced time $\tau=t\sqrt{m\sigma^2\over\epsilon}$, hence 0.005$\tau$ time step is used in the simulation. The system is initially heated up to a high temperature of 1.2 in the unit of the reduced   temperature, $T^*={k_B T\over\epsilon}$, where the system is  in a liquid state. After a quick equilibration, the system is rapidly cooled down to $T^*$ = 0.3 or 0.2 at a cooling rate of 3.3 x $10^{10}$ K/s and then equilibrated again for another two million time steps. In the production run, the system is simulated for three million time steps, saving configurations at every 100 time steps or 0.05$\tau$. The average reduced density of the system is $\rho^*=\rho\sigma^3$= 1.17 or 1.19. 

In the liquid-solid nucleation test case, the WCNs of particles are generated from the configuration data with normalized Gaussian weighting functions in combination with $g(r)$ of the system. However, for supercooled binary LJ liquid configurations( A and B particles are treated as the same), the local structures of particles captured by their WCNs are much more complicated and noisy in contrast to the solid-liquid nucleation case  where solid and liquid-like particles are sharply different from each other due to their distinctly different symmetries. Thus, an extra step is implemented to average out the WCNs of neighboring particles to remove some noises, namely, $\overline{WCN}_i={\sum_{i \neq j}^{N_b}WCN_j}/{N_b}$,
where $N_b$ is the number of neighboring particles in each shell.

\begin{figure}
\resizebox{0.8\columnwidth}{!}
{
\includegraphics[width=1.5in]{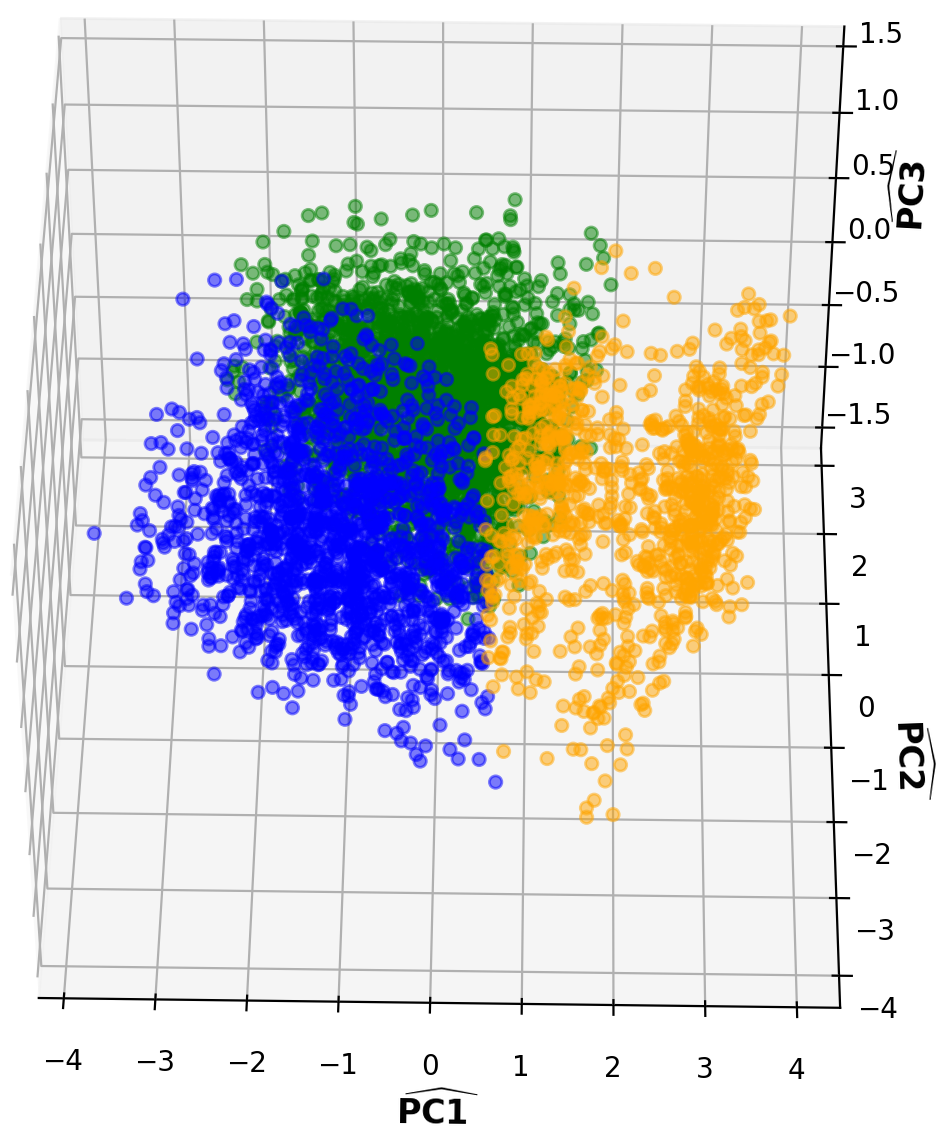}
\includegraphics[width=1.5in]{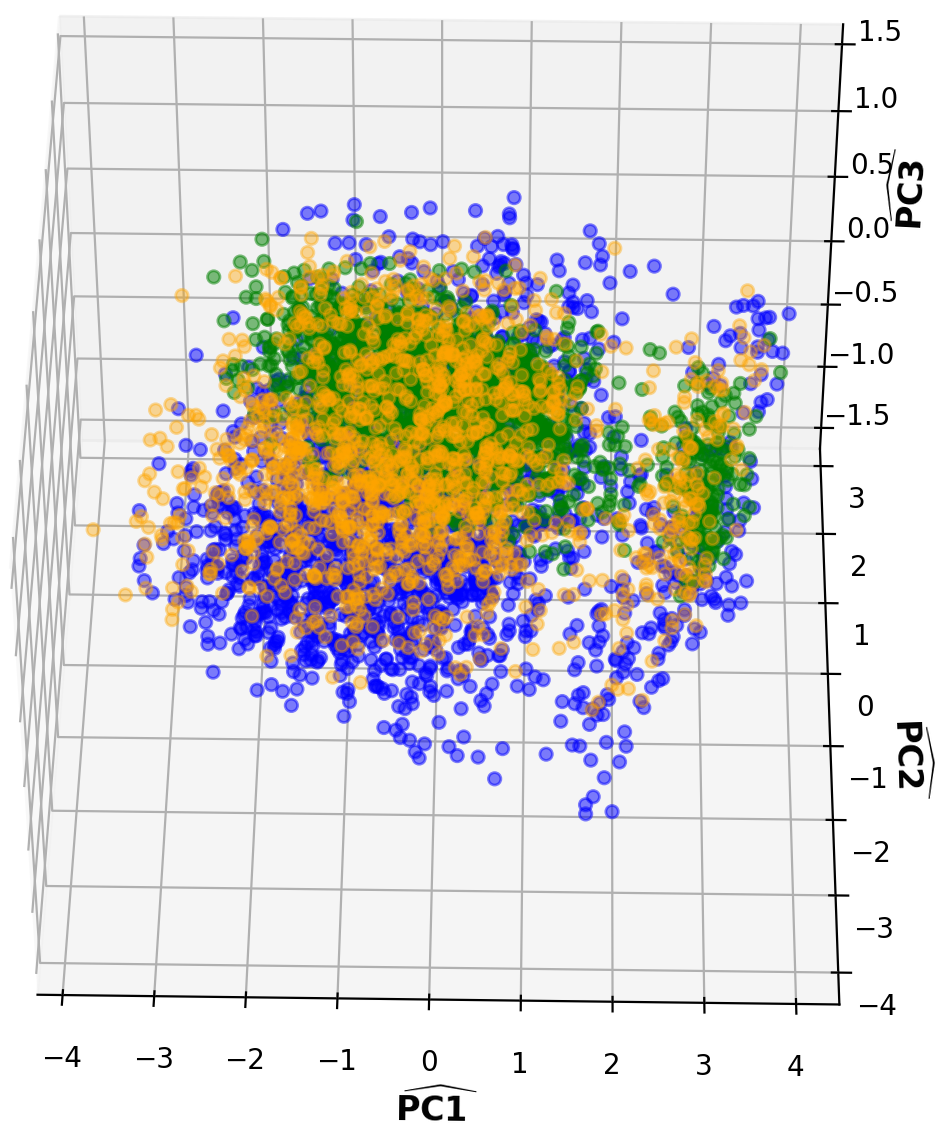}
}
\resizebox{0.8\columnwidth}{!}
{
\includegraphics[width=1.5in]{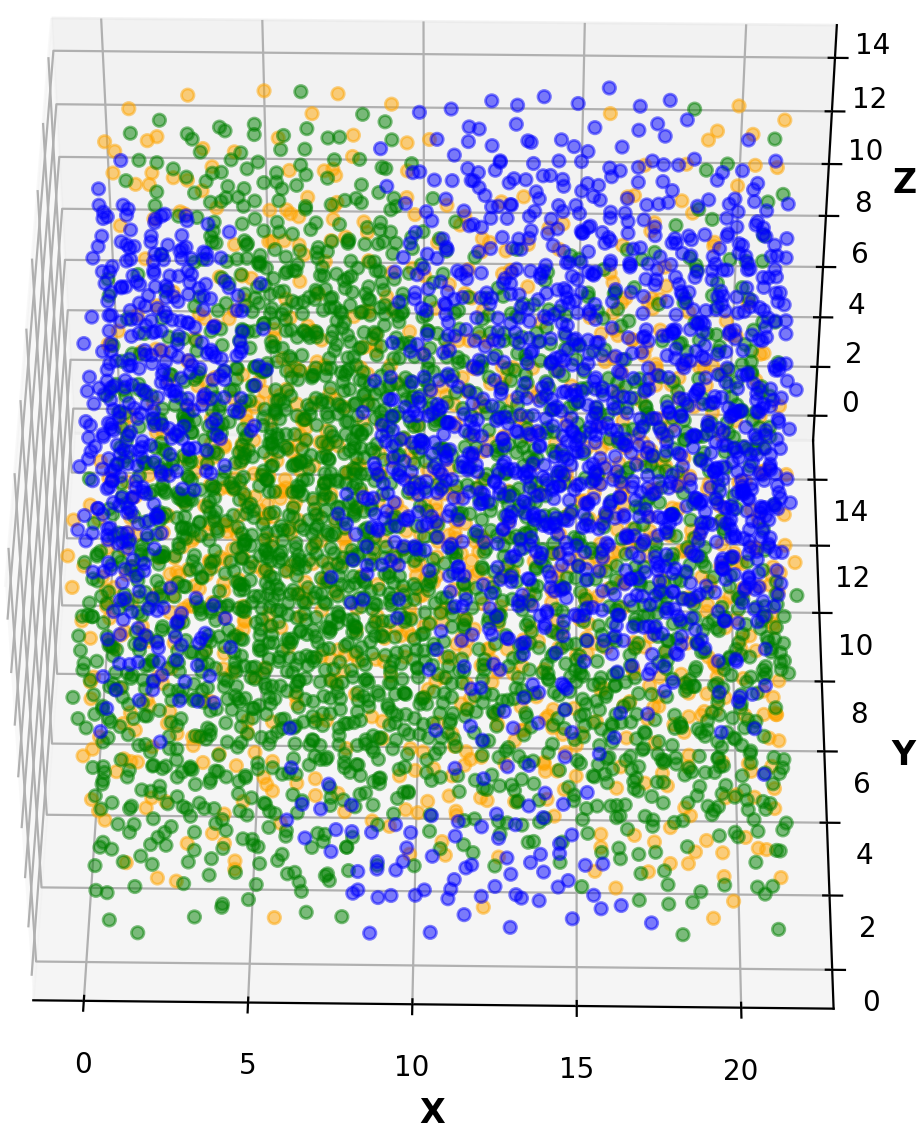}
\includegraphics[width=1.5in]{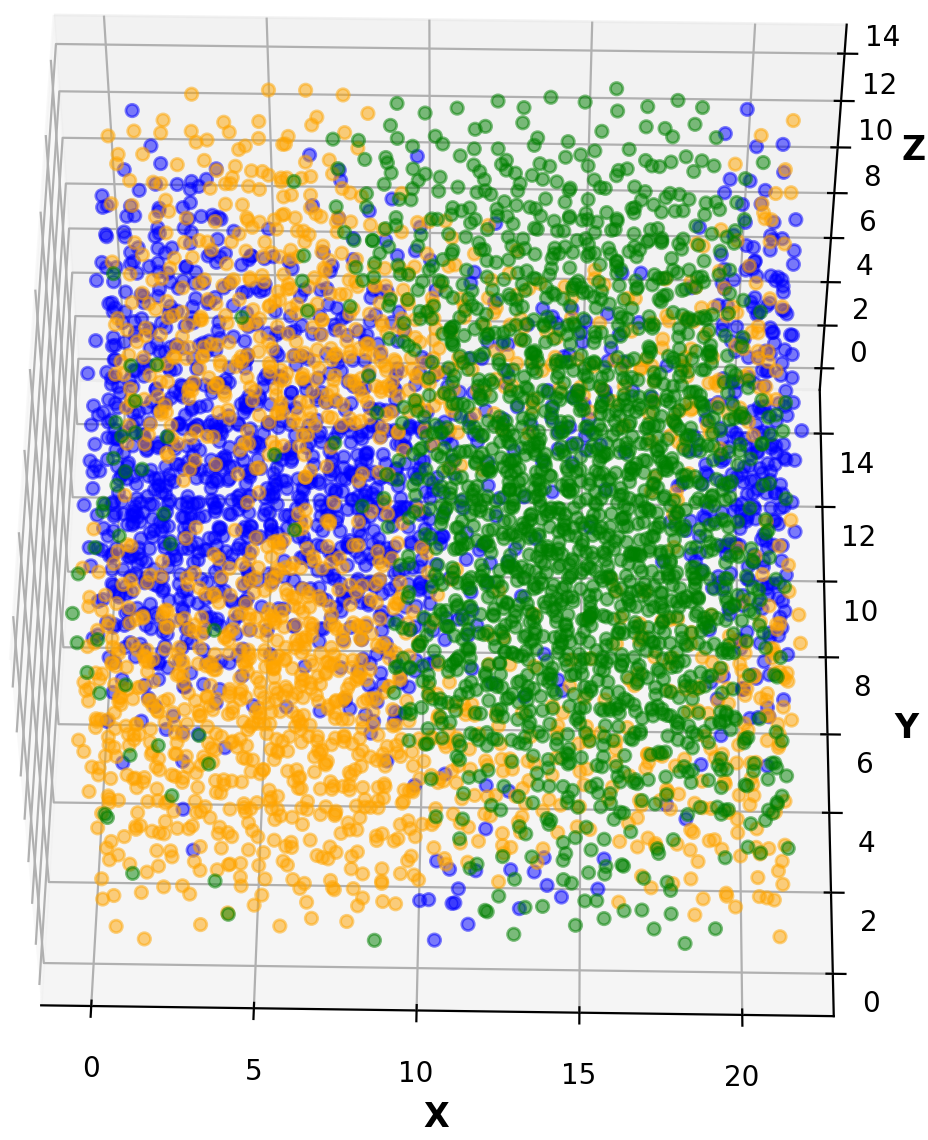}
}
\parbox{6.3in}
{
\caption{ A 3D snapshot of 3 meso-states of 5000 particles system at $T^*=0.3$, cooling rate at 3.3 x 10$^{10}$ K/s for  both direct mapping and co-learning method. Panels a) and c) are for the PC-space and configurational space from direct mapping while panels b) and d) presents clusters formed in the PC-space and configurational space after convergence from co-learning. The hat symbol for labelling axes of the PC space represents the inner product of a particle's $\overline{WCN}$s with the PCs basis, in this case for the first three PC components. Blue, orange and green-colored particles belong to meso-state 1, meso-state 2 and meso-state 3, respectively.}  
\label{fig:nano-domain}
}
\end{figure}

The $\overline{WCN}$s of particles are used for PCA, and then a K-means clustering with an Elbow convergent test.  The Elbow test  suggests that K being either 3 or 4 depend the temperature and cooling rate for the cases studied. Naturally 
a better implementation might be some Bayesian based algorithms, but the K-means clustering serves our purpose in this study.  Given a choice of K=3, K-means clustering is used to classify the system in the PC-space based on its structural features into 3 meso-states in Fig. \ref{fig:nano-domain}a, then the clustering in the configurational space can be obtained  from the identity of the particles in a meso-state.  However, a direct transfer mapping from the PC-space to the configurational space shows some uncertainties in assignments such as orange particles are dispersed in the Fig. \ref{fig:nano-domain}c. 

This implies that the transferring information from one space to another is highly non-trivial, thus a co-learning strategy is proposed: i) perform K-means clustering in the PC-space;
ii) use the initial knowledge of the clustering  from the PC-space to perform a K-means clustering in the configurational space; iii) use the clustering knowledge from the configurational space to perform a Gaussian Mixture (GM) classification in the PC-space to soft the hard assignment from K-means; iv) perform GM in the configurational space from the clustering knowledge from the PC-space; v) iteratively perform GM classification in both spaces until convergence.  

This co-learning strategy is highly efficient because it expands and complements the missing information from a direct transfer from the PC-space to the configurational space. When the  iterations converged (normally after 3 or 4 iterations), this co-learning strategy yields exceptionally clean meso-states in the PC-space as shown Fig. \ref{fig:nano-domain}b  and nano-domains in the configurational space as displayed in Fig. \ref{fig:nano-domain}d , which represent unambiguously aggregated clusters formed in both PC-space and configurational space at a simulation snapshot of $T^*=0.3$. These cluster identities remain unchanged after few iterations and present clear evidence of spatial heterogeneity of the system at a supercooled condition. 

\begin{figure}
\resizebox{0.8\columnwidth}{!}
{
\includegraphics[width=1.5in]{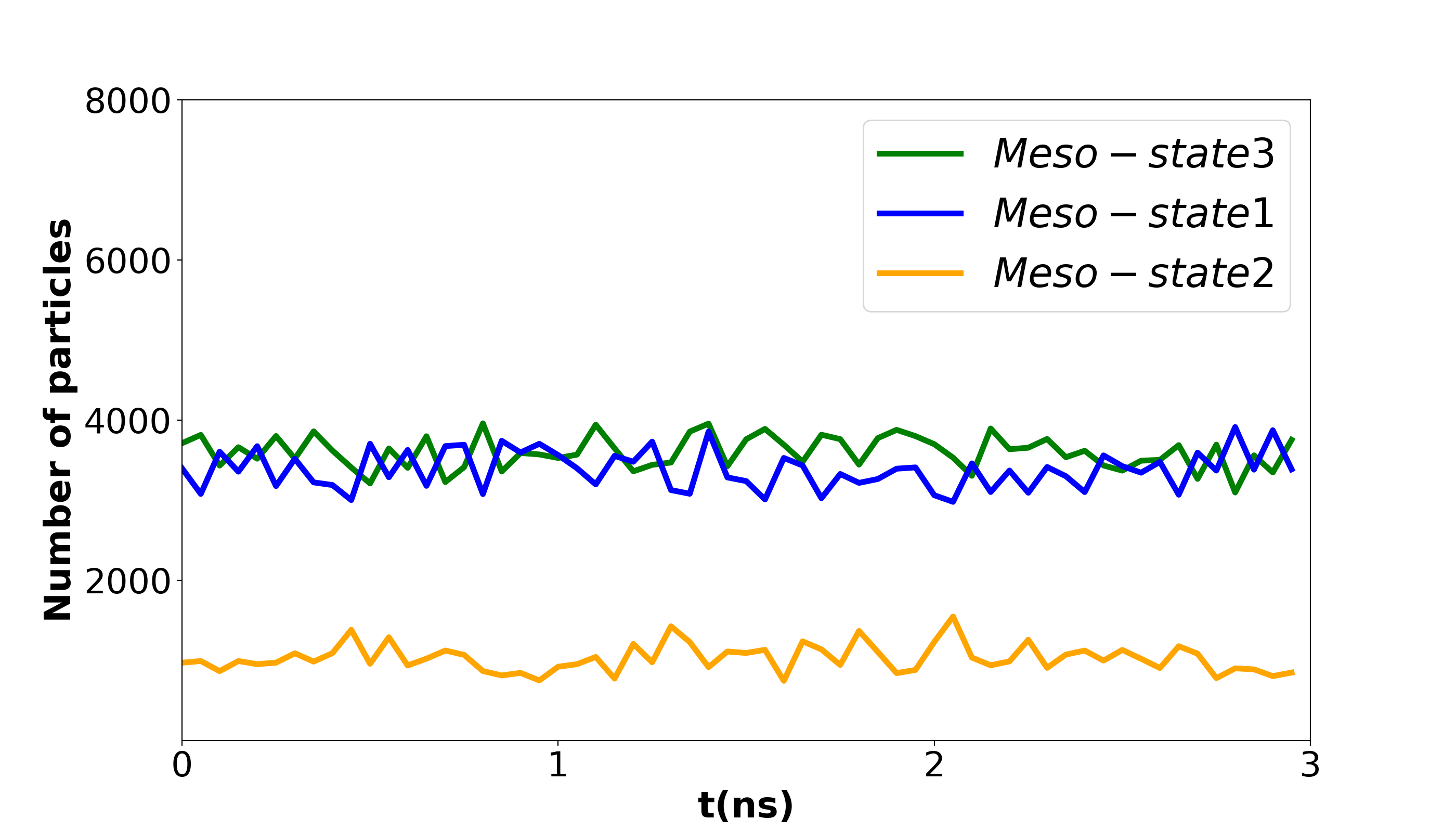}
\includegraphics[width=1.5in]{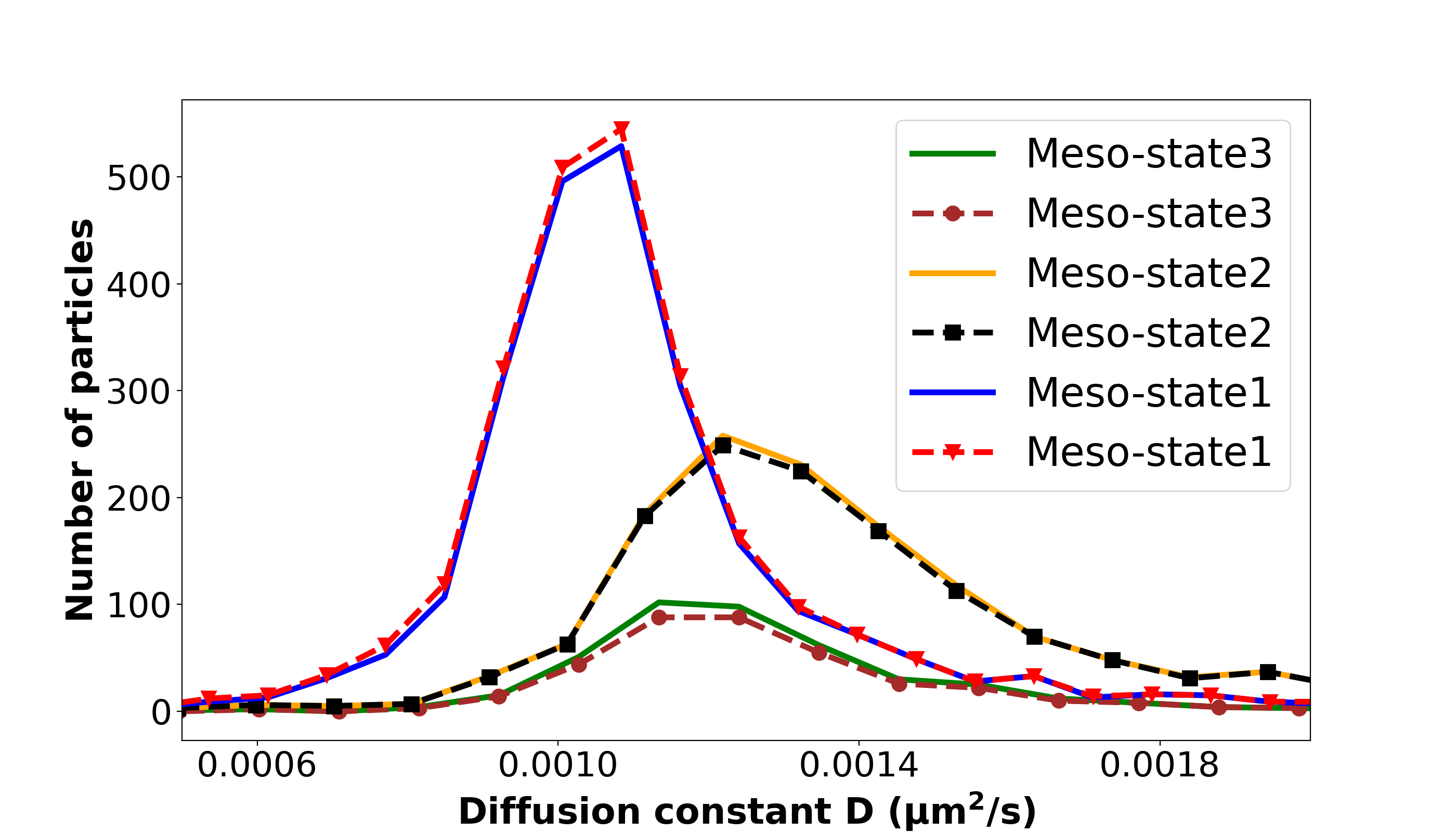}
}
\parbox{6.3in}
{
\caption{ A number of particle fluctuations and dynamics of 3 meso-states of 16000 particles system at $T^*=0.3$, cooling rate is 3.3x10$^{10}$ K/s. Panel a) illustrates the particles fluctuations over 3ns while panel b) show the distribution of diffusion constants of core particles within a single patch of each meso-state over 1ns.}
\label{fig:fluc-dynamics}
}
\end{figure}

Using this co-learning strategy, the lifetime of these nano-domains can be evaluated. Figure.~\ref{fig:fluc-dynamics}a displays fluctuations of particles number for a single patch of each meso-states over a few nano-seconds.

Furthermore, as the system size increases, to have a few meso-states to capture the structural heterogeneity, but in configurational space, we would expect that  a meso-state in the PC-space should be able to bifurcate into small domains or patches which roughly have the same size in the configurational space to tile up the whole space. This is exactly what we observed from the implementation of the co-learning strategy with increasing system sizes shown in Table~\ref{tab:bifurcation}. As the system size grows, each meso-state split into different domains with a consistent size. The larger system is, the higher number of patches each meso-state owns. 

\begin{table}
\parbox{3.3in}
{
	\caption {Number of particles of various patches belong to different meso-states for two particle system sizes at $T^*=0.3$, cooling rate at 3.3x10$^{10}$ K/s}
	\label{tab:bifurcation}
}
\resizebox{\columnwidth}{!}
{
  	\begin{tabular}{ | cc | c | c | c | c | c | c | c | c}
\cline{1-9}
& & \textbf{Patch 1} & \textbf{Patch 2} & \textbf{Patch 3} & \textbf{Patch 4} & \textbf{Patch 5} & \textbf{Patch 6} & \textbf{Patch 7} \\ \cline{1-9}

\multicolumn{1}{ |c  }{\multirow{2}{*}{\textbf{16000 particles }} } &
\multicolumn{1}{ |c| }{\textbf{Meso-state 1}} & 3519 & 3528  &   &   &   &   &   &   \\ \cline{2-9}
\multicolumn{1}{ |c  }{}                        &
\multicolumn{1}{ |c| } {\textbf{Meso-state 3}} & 3564 & 3558 &   &   &   &   &   &    \\ \cline{2-9}
\multicolumn{1}{ |c  }{}                        &
\multicolumn{1}{ |c| }{\textbf{Meso-state 2}} & 910 & 921 &   &   &   &   &   &    \\ \cline{1-9}
\multicolumn{1}{ |c  }{\multirow{2}{*}{\textbf{50000 particles }} } &
\multicolumn{1}{ |c| }{\textbf{Meso-state 1}} & 3355 & 3277 & 3298 & 3432  & 3487  & 3469 & 3541 \\ \cline{2-9}
\multicolumn{1}{ |c  }{}                        &
\multicolumn{1}{ |c| }{\textbf{Meso-state 3}} & 3619 & 3773 & 3717 & 3720  & 3733 & 3613 & \\ \cline{2-9}
\multicolumn{1}{ |c  }{}                        &
\multicolumn{1}{ |c| }{\textbf{Meso-state 2}} & 927 & 957 & 1002 & 1080  &   &   &   & \\ \cline{1-9}
	\end{tabular}
}
\end{table}

Finally, armed with the configurational identity of particles in various nano-domains,  the difference in the distribution of diffusion constants of each meso-state in Figure.~\ref{fig:fluc-dynamics}b) confirms the heterogeneous dynamics of the system. 

At $T^*$=0.2, the same analysis presents a consistent picture as shown for the $T^*$=0.3 but with much slower dynmaics.

In summary, the results reported in this study indicate that  a practical co-learning strategy can be developed  to study the  existence of structural and dynamic heterogeneities in a supercooled liquid. In our approach, ML algorithms such as PCA with K-means and GM clustering  provide a robust method of classifying particles to form nano-domains in both structural and configurational spaces. Dynamics of such meso-states are different from the distribution of diffusion constants of cores particles. To characterize such spatial and dynamical heterogeneities, meso-states, a representation of structural heterogeneity, show  structural stability over many nanoseconds simulations. To demonstrate the structural integrity of the meso-states, it is shown that each meso-states will split into multiple patches with equal size to tile up the whole simulation box. Furthermore, the size of the patches of a meso-state remains the same regardless of the increasing particle size of the system. Therefore, our classification scheme paves the way for further statistical mechanics analysis of supercooled liquids.

\section{Acknowledgement}

This work is supported by the Division of Chemical and Biological Sciences, Office of Basic Energy Sciences, U.S. Department of Energy, under Contact No. DE-AC02-07CH11358 with Iowa State University. 
\bibliographystyle{apsrev4-1}
\bibliography{nano_domain} 

\end{document}